\documentclass[reprint,amsmath,amssymb, notitlepage,prb]{revtex4-1}
\usepackage{graphicx}
\usepackage{dcolumn}
\usepackage{bm}
\usepackage{physics}

\allowdisplaybreaks[1]

\newcommand{\average}[1]{\ensuremath{\langle#1\rangle} }

\begin{document}
	
	\title{Acoustic spin current generation in superconductors}
	\author{Takumi Funato}
	\affiliation{Center for Spintronics Research Network, Keio University, Yokohama 223-8522, Japan}
	\affiliation{%
		Kavli Institute for Theoretical Sciences, University of Chinese Academy of Sciences, Beijing, 100190, China.
	}
	\author{Ai Yamakage}
	\affiliation{
		Department of Physics, Nagoya University, Nagoya 464-8602, Japan
	}
	\author{Mamoru Matsuo}
	\affiliation{%
		Kavli Institute for Theoretical Sciences, University of Chinese Academy of Sciences, Beijing, 100190, China.
	}%
	\affiliation{%
		CAS Center for Excellence in Topological Quantum Computation, University of Chinese Academy of Sciences, Beijing 100190, China
	}%
	\affiliation{
		Advanced Science Research Center, Japan Atomic Energy Agency, Tokai, 319-1195, Japan
	}
	\affiliation{RIKEN Center for Emergent Matter Science (CEMS), Wako, Saitama 351-0198, Japan}

	\date{\today}
	
	\begin{abstract}
	We theoretically study the generation of spin current due to a surface acoustic wave (SAW) in a superconductor.
	We model an s-wave superconductor as the mean-field Hamiltonian and calculate spin current generated via spin-vorticity coupling based on kinetic theory.
	The results suggest that the spin current can be driven in a single superconductor layer, and our estimation suggests that the detectable magnitude of the spin current can be generated in aluminum.
	Our proposal may contribute to the advancement of spin transport in superconductors from application and fundamental physics aspects.
	\end{abstract}

	\pacs{Valid PACS appear here}
	\maketitle
	
	\section{introduction}

	Spin-related phenomena in hybrid systems comprised of superconductors and ferromagnetic metals have been studied for several decades\cite{zutic2004Spintronics,linder2015Superconducting,beckmannSpinManipulationNanoscale2016,bergeretColloquiumNonequilibriumEffects2018,quayOutofequilibriumSpinTransport2018,han2020Spin,ohnishiSpintransportSuperconductors2020}. 
	The relation between spin transport and phenomena specific to superconductors, such as the proximity effect and Andreev reflection, has primarily been investigated since the superconducting coherent length scale is the same as the spin relaxation length scale\cite{beckmann2004Evidence,russo2005Experimental,cadden-zimansky2006Nonlocal,cadden-zimansky2007Charge,yeyati2007Entangled,kleine2009Contact,kleine2010Magnetic}. 
	Tunneling resistance of the superconducting point contact provides the spin polarization of a ferromagnetic metal\cite{soulen1998Measuring}. 
	An oscillatory superconducting transition temperature which depends on the magnetization and thickness of the adjacent ferromagnetic metal layer has also observed\cite{jiang1995Oscillatory,mercaldo1996Superconductingcriticaltemperature,muhge1996Possible,fominov2001Critical,garifullin2002Reentrant,tagirov2002Reentrant,maekawa2002Spin,yamashita2002Spina,takahashi2003Spin,morten2004Spin}. 
	The crossed Andreev reflection and quasiparticle tunneling transport has been investigated in trilayer systems comprised of ferromagnetic metal/superconductor/ferromagnetic metal\cite{beckmann2004Evidence,russo2005Experimental,cadden-zimansky2006Nonlocal,cadden-zimansky2007Charge,yeyati2007Entangled,kleine2009Contact,kleine2010Magnetic,ohnishi2015Significant}. 
	An extremely long quasiparticle spin transport in aluminum (Al) thin films embedded in a magnesium oxide (MgO) insulating layer has also been observed\cite{yang2010Extremely}. 
	Injecting a pure spin current into a superconductor with non-local spin valve systems allows for studying spin-related phenomena without the influence of charge transport\cite{miura2006Non,poli2008Spin,ohnishi2010Nonlocal,quaySpinImbalanceSpincharge2013,wakamura2014Spin,wakamura2015Quasiparticlemediated,takahashi2008Spin,takahashi2011Spin}. 
	Characteristic spin pumping in a bilayer system composed of superconductor and ferromagnetic insulator is typically studied based on microscopic theory\cite{inoue2017Spin,kato2019Microscopic,ominatoAnisotropicSuperconductingSpin2021,ominato2022Ferromagneticb}.
	In a recent related study, the negative resistance state in niobium diselenide (NbSe$_2$) induced by surface acoustic waves was reported\cite{yokoi2020Negative} and the superconducting diode effect in a [Nb/V/Ta]$_n$ superlattice with broken spatial inversion symmetry was observed\cite{ando2020Observation,miyasaka2021Observation,daido2022Intrinsic}.
	A system combining a superconductor and a ferromagnetic metal is a good probe for investigating spin transport in superconductors.
	
    Here, we propose spin-current generation by a surface acoustic wave (SAW) in a single superconductor layer.
    In the conventional method of spin transport in a superconductor, spin injection from an adjacent ferromagnetic material is necessary.
    One reason for this is that manipulation of spin current by both external electric and magnetic fields is difficult in a superconductor.
	On the other hand, mechanical means can be used to drive spin current in a superconductor, without such restrictions, since the vorticity associated with the mechanical motion is coupled to the spin angular momentum of the electrons, not the magnetic moment.
	Mechanical spin-current generation in a superconductor remains an open problem requiring study.

	Mechanical spin-current generation based on the conversion of angular momentum from mechanical rotation into electron spin has attracted much attention in spintronics.
	The underlying origin is proposed to be spin-vorticity coupling (SVC); the coupling between electron spin and the effective field associated with mechanical rotation.
    The SVC-mediated mechanism has been experimentally confirmed.
    It has been presented that spin current is generated by the vorticity of liquid metal laminar flow and spin Hall voltages were observed\cite{takahashi2016Spin,takahashi2020Giant,tokoro2022Spin,tabaeikazerooni2020Electron,tabaeikazerooni2021Electrical}. 
    It was also reported that spin current generated by local lattice rotation associated with a SAW was observed through spin-wave resonance\cite{kobayashi2017Spin,kurimune2020Observation,tateno2020Electrical}. 
    According to this result, spin current is successfully generated in copper (Cu) with weak spin-orbit interaction, which is essential in conventional spin-current generation.
    SVC can broaden the range of materials capable of spin current generation because of its universal effect.

	In this paper, we study generation of spin current by a SAW in an s-wave superconductor.
	As a model, we consider the s-wave superconductor to which a SAW is applied and calculate the spin current generated via SVC based on kinetic theory up to the first order in vorticity.
	We demonstrate spin current generated by a Rayleigh-type SAW in a single superconducting layer and estimate the driven spin current.
	We expect that our proposal will contribute to the development of spin transport in superconductors.

	\section{Model}
	We consider an s-wave superconductor in the presence of normal and spin-orbit impurities.
	The total Hamiltonian is given by
	\begin{align}
	    H = H_{\text{sc}} + H_{\text{imp}} + H_{\text{so}}.
	    \label{total_ham}
	\end{align}
	The first term in Eq.~(\ref{total_ham}) is the mean-field Hamiltonian which describes the s-wave superconductor and is given by
	\begin{align}
	    H_{\text{sc}} = \frac{1}{2} \sum_{\bm k} \Phi_{\bm k}^{\dagger} \begin{pmatrix}
	    \xi_{k} & \Delta i\sigma_y \\[2pt]
	    -\Delta i\sigma_y & -\xi_{k}
	    \end{pmatrix}
	    \Phi_{\bm k},
	    \label{total_hamiltonian}
	\end{align}
	where $\Phi_{\bm k}=(c_{\bm k\uparrow},c_{\bm k\downarrow},c^{\dagger}_{-\bm k\uparrow},c^{\dagger}_{-\bm k\downarrow})$ is the four-component Nambu spinor with $c_{\bm k\sigma}^{\dagger}(c_{\bm k\sigma})$ being the creation (annihilation) operator of the spin $\sigma$ electrons, $\xi_k=k^2/2m-\mu$ is the energy of conduction electrons measured from the chemical potential $\mu$, and $\sigma_y$ is the $y$-component of the Pauli matrix.
	$\Delta$ is the superconducting energy gap of Bardeen--Cooper--Schrieffer (BCS) theory, determined by the gap equation:
	\begin{align}
	    \text{ln}\left( \frac{T}{T_c} \right) \Delta = 2\pi T \sum_{\epsilon_n} \left( \frac{\Delta}{\sqrt{\epsilon_n^2+\Delta^2}} - \frac{\Delta}{\epsilon_n} \right),
	\end{align}
	where $\epsilon_n=(2n+1)\pi T$ is the Matsubara frequency and $T_c$ is the superconducting transition temperature.
	The phenomenological temperature dependence of the superconducting energy gap is assumed to be
	\begin{align}
        \Delta(T) \simeq 1.76 k_B T_c \tanh \left( 1.74 \sqrt{\dfrac{T_c}{T}-1} \right),
    \end{align}
    
    The second and third terms in Eq.~(\ref{total_hamiltonian}) describe coupling to the impurity potential and impurity spin-orbit interaction, respectively:
    \begin{align}
        H_{\text{imp}} &= \frac{1}{2}\sum_{\bm k\bm k'} \Phi^{\dagger}_{\bm k'}
        \begin{pmatrix}
            V_{\bm k'-\bm k} & 0 \\
            0 & -V_{\bm k'-\bm k}^*
        \end{pmatrix}
        \Phi_{\bm k},
        \\
        H_{\text{so}} &= \frac{i\lambda_{\text{so}}}{2} \sum_{\bm k\bm k'} (\bm k'\times \bm k) \cdot \Phi^{\dagger}_{\bm k'}
        \begin{pmatrix}
            V_{\bm k'-\bm k}\bm \sigma & 0 \\
            0 & V_{\bm k'-\bm k}^* \bm \sigma^*
        \end{pmatrix}
        \Phi_{\bm k},
    \end{align}
     where $V_{\bm k'-\bm k}$ is the Fourier component of the impurity potential $V_{\text{imp}}(\bm r)$, $\lambda_{\text{so}}$ is the strength of spin-orbit interaction, $\bm \sigma =(\sigma_x,\sigma_y,\sigma_z)$ are the Pauli matrices in spin space.
    In this paper, we assume a short-range impurity potential, i.e., $V_{\text{imp}}=u_i\sum_j\delta (\bm r-\bm r_j)$, where $u_i$ is the strength of the impurity potential, $\bm r_j$ is the position of the $j$-th impurity.
    Assuming a uniformly random distribution of the impurities, we average over the impurity positions as $\average{V_{\bm k}V_{\bm k'}}_{\text{imp}} = n_iu_i^2 \delta_{\bm k+\bm k',\bm 0} + n_i^2u_i^2 \delta_{\bm k,\bm 0}\delta_{\bm k',\bm 0}$.
    
    When a SAW with the frequency $\omega$ and wavenumber $\bm q$ is applied into the s-wave superconductor, the electron spins are coupling to the lattice rotational motion via SVC.
    The $z$-axis is chosen to be parallel to the vorticity associated with the SAW, and the SVC Hamiltonian is described by
	\begin{align}
	    H_{\text{sv}} = -\frac{\hbar}{8}\sum_{\bm k\sigma}
	    \Phi^{\dagger}_{\bm k_+} 
	    \begin{pmatrix}
	        \sigma_z & 0\\
	        0 & -\sigma_z
	    \end{pmatrix} 
	    \Phi_{\bm k_-} \Omega(\bm q,\omega) e^{-i\omega t},
	\end{align}
	where $\bm k_{\pm}=\bm k\pm \bm q/2$ and $\bm \Omega(\bm q,\omega)$ is the Fourier components of the the vorticity of the lattice $\bm \Omega(\bm r,t) = \nabla \times \partial_t \bm u(\bm r,t)$ with the lattice velocity field $\bm u(\bm r,t)$.

    The $z$-polarized spin-current operator is given by
	\begin{align}
	    \bm j_{\text s}(\bm q) = \frac{1}{2} \sum_{\bm k\sigma} 
	    \frac{\hbar \bm k}{m} \Phi_{\bm k_-}^{\dagger}
	    \begin{pmatrix}
	        \sigma_z & 0\\
	        0 & \sigma_z
	    \end{pmatrix}
	    \Phi_{\bm k_+}.
	\end{align}
	Note that the anomalous velocity due to impurity spin-orbit interaction is negligible because it do not contribute the spin-current generation in this setup.

	To diagonalize the mean-field Hamiltonian, we perform the Bogoliubov transformation:
	\begin{gather}
	    \begin{pmatrix}
	        c_{\bm k} \\ c^{\dagger}_{-\bm k}
	    \end{pmatrix}
	    = \begin{pmatrix}
	        u_k & v_k^*i\sigma_y \\
	        -v_ki\sigma_y & u_k^*
	    \end{pmatrix}
	    \begin{pmatrix}
	        \gamma_{\bm k} \\ \gamma^{\dagger}_{-\bm k}
	    \end{pmatrix}
	    ,
	\end{gather}
	where $|u_k|^2=(1+\xi_k/E_k)/2$ and $|v_k|^2=(1-\xi_k/E_k)/2$ are the coherent factors with the quasi-particle energy dispersion $E_k=\sqrt{\xi_k^2+\Delta^2}$, and $\gamma_{\bm k}=(\gamma_{\bm k\uparrow},\gamma_{\bm k\downarrow})$ and $\gamma^{\dagger}_{-\bm k}=(\gamma^{\dagger}_{-\bm k\uparrow},\gamma^{\dagger}_{-\bm k\downarrow})$ are the creation and annihilation operators of the quasiparticles, respectively.
	The mean-field and spin-vorticity coupling Hamiltonians are expressed as
	\begin{gather}
	    H_{\text{sc}} = \sum_{\bm k} E_k \gamma_{\bm k}^{\dagger} \gamma_{\bm k},
	    \\
	    H_{\text{sv}} = -\frac{\hbar}{4}\sum_{\bm k,\bm q}
	     \gamma_{\bm k_+}^{\dagger} \sigma_z \gamma_{\bm k_-}
	    \Omega(\bm q,t),
	\end{gather}
	where $\gamma_{\bm k} = (\gamma_{\bm k\uparrow},\gamma_{\bm k\downarrow}) $.
	The spin-current operator is given by
	\begin{align}
	    \bm j_{\text s}(\bm q)
	   &=  \sum_{\bm k} \bm v_{\bm k} 
	  \gamma_{\bm k_-}^{\dagger} \sigma_z \gamma_{\bm k_+}
	  ,
	\end{align}
    where $\bm v_{\bm k}=\partial E_{k}/\partial \hbar \bm k=(\hbar \bm k/m)(\xi_k/E_k)$ is the velocity of the quasiparticle.

    \section{Formulation}
    The expectation value of the spin current is given by 
    \begin{align}
	    \average{\bm j_{\text{s}}(\bm r,t)} = ie \lim_{\tau \rightarrow 0} \sum_{\bm k\sigma,\bm q} e^{i\bm q\cdot \bm r} \sigma \bm v_{\bm k} 
	    g^<_{\sigma}(\bm k_+t_+;\bm k_-t_-)
	    ,
	\end{align}
	where $t_{\pm}=t\pm \tau/2$ and $\sigma$ represents the quasiparticle spin with $+$ for up spin and $-$ for down spin.	$g^<_{\sigma}(\bm k_+t_+;\bm k_-t_-)$ is the lesser component of the nonequilibrium Green function for the spin $\sigma$ quasiparticle:
	\begin{align}
	   g_{\sigma}(\bm k_+t_+;\bm k_-t_-)
	   = -i \average{T_C \gamma_{\bm k_+\sigma}(t_+) \gamma_{\bm k_-\sigma}^{\dagger}(t_-)},
	\end{align}
	where $\gamma_{\bm k\sigma}(t)$ is the Heisenberg representation of the quasiparticle annihilation operator, $T_C$ is the path-ordered operator for the Keldysh contour, and $\average{\cdots}=\text{tr}(\hat \rho \cdots)$ is the expectation value the density operator $\hat \rho$.
	
	We introduce the Wigner function obtained by Fourier transforming the lesser component of the nonequilibrium Green function with respect to the relative coordinates and time $\bm \rho$ and $\tau$, respectively:
	\begin{align}
	    \phi^w_{\bm k\epsilon,\sigma}(\bm r,t) = -i \int d^3\rho d\tau e^{-i(\bm k \cdot \bm \rho -\epsilon \tau)} g^<_{\sigma}(\bm r_+t_+;\bm r_-t_-),
	\end{align}
	where we define $\bm r_{\pm} = \bm r\pm \bm \rho/2$ with barycentric time and coordinates $t=(t_++t_-)/2$ and $\bm r=(\bm r_++\bm r_-)/2$, respectively.
	The spin current can be expressed by the Wigner function:
	\begin{align}
	    \average{\bm j_s(\bm r,t)} = -e \sum_{\bm k\sigma} \sigma \bm v_{\bm k} \int \frac{d\epsilon}{2\pi} \phi^w_{\bm k\epsilon,\sigma}(\bm r,t).
	\end{align}
	The Wigner distribution function is governed by the Kadanoff--Baym equation treating impurity scattering and impurity spin-orbit scattering in a perturbative manner:
	\begin{align}
	    &\left( \partial_t + \bm v_{\bm k} \cdot \nabla + \sigma \frac{\hbar}{4} (\nabla_x \Omega_{\bm rt}) \cdot \nabla_p \right) \phi^w_{\bm k\epsilon,\sigma} -\{ \text{Re}\,\Sigma^R, \phi^w_{\bm k\epsilon,\sigma} \}  \nonumber \\ 
	    &  - \{ i\Sigma^<,\text{Re}\, g^R_{\bm k\epsilon,\sigma}\}
	   = g^>_{\bm k\epsilon,\sigma} \Sigma^< - \Sigma^>g^<_{\bm k\epsilon,\sigma},
	\end{align}
	where $\nabla_x = (\partial_t, \nabla)$ and $\nabla_p=(-\partial_{\epsilon},\partial_{\bm k})$ are the derivative of the four-vectors, $\{ A,B\} = \nabla_x A \cdot \nabla_p B - \nabla _p A \cdot \nabla_x B$ is the Poisson bracket, and 
	$\Sigma$ is the self-energy due to impurity scattering and impurity spin-orbit scattering.
	
	We assume that the spectrum function has a delta-function peak, and the Wigner distribution function is given by $\phi^w_{\bm k\epsilon}(\bm r,t)=2\pi \delta(\epsilon -E_k) f_{\bm k\sigma}(\bm r,t)$.
	Here, $f_{\bm k\sigma}(\bm r,t)$ is the distribution function, defined by 
	\begin{align}
	    f_{\bm k\sigma}(\bm r,t) = \int \frac{d\epsilon}{2\pi} \phi^w_{\bm k\epsilon,\sigma}(\bm r,t).
	\end{align}
	The expectation value of the spin current can be given by
	\begin{align}
	    \average{\bm j_s(\bm r,t)} = -e\sum_{\bm k} \bm v_{\bm k} \Bigl[f_{\bm k\uparrow}(\bm r,t)-f_{\bm k\downarrow}(\bm r,t)\Bigr].
	    \label{spin_curent_exp}
	\end{align}
	Integrating both sides of the Kadanoff--Baym equation with respect to the energy $\epsilon$, we derive the Boltzmann equation, which governs the distribution function, as:
    \begin{align}
	    &\frac{\partial f_{\bm k\sigma}}{\partial t}  + \bm v_{\bm k} \cdot \frac{\partial f_{\bm k\sigma}}{\partial \bm r} +\bm F_{\sigma} \cdot \frac{\partial f_{\bm k\sigma}}{\partial \hbar\bm k}
	    = I_{\bm k\sigma}[f],
	    \label{beq1}
	\end{align}
	where $\bm F_{\sigma}$ is the spin-dependent force due to the SVC, given by
	\begin{align}
	    \bm F_{\sigma} =  \sigma \frac{\hbar}{4}\frac{\partial \Omega(\bm r,t)}{\partial \bm r},
	    \label{svc_force}
	\end{align}
	and $I_{\bm k\sigma}[f]$ is the collision term, given by
	\begin{align}
	    I_{\bm k\sigma}[f] = \int \frac{d\epsilon}{2\pi} (g^>_{\bm k\epsilon,\sigma} \Sigma^< - \Sigma^>g^<_{\bm k\epsilon,\sigma}). 
	\end{align}
	Calculating the self-energy in the second order Born approximation, the collision term up to second order in the spin-orbit interaction is derived as
	\begin{align}
	    I_{\bm k\sigma}[f] = -\frac{f_{\bm k\sigma}-f_{k\sigma}^{\text{eq}}}{\tau_k} - \left. \frac{f_{\bm k\sigma}-f_{k'-\sigma}^{\text{eq}}}{\tau_{s,k}}\right|_{E_{k\sigma}=E_{k'-\sigma}},
	\end{align}
	where $f^{\text{eq}}_{k\sigma}$ is the local equilibrium distribution function for the quasiparticles, $E_{k\sigma}=E_k - \sigma \hbar \Omega/4$ is the quasi-particle energy including the SVC,
	$\tau_k = \tau E_k/|\xi_k|$ is the momentum-scattering time, and $\tau_{s,k}=6\tau_k /\lambda_{\text{so}}^2k_F^4 (2+\cos ^2\theta)$ is the spin-flip scattering time with $\tau ^{-1} = (2\pi /\hbar) n_iu_i^2 N(\mu) (1+\frac{2}{3}\lambda_{\text{so}}^2k_F^4)$.

    \section{Calculation}
    First, let us solve the Boltzmann transport equation.
    We assume the temporal and spatial variations of the vorticity $\Omega$ are much slower than the relaxation time $\tau$ and the mean-free path $l$ of the quasiparticles, respectively.
    The deviation of the nonequilibrium state fluctuated by the SVC is then approximately characterized by the local equilibrium distribution function, given by
    \begin{align}
	    f_{k\sigma}^{\text{eq}} = f_0(E_{k\sigma}-\sigma \delta \mu_s),
	\end{align}
	$\delta \mu_s=(\mu_{\uparrow}-\mu_{\downarrow})/2$ is the spin accumulation with $\mu_{\uparrow}$($\mu_{\downarrow}$) being the chemical potentials of up-spin (down-spin) quasiparticles.
	The local equilibrium distribution function $f_{k\sigma}$ can be expanded as follows:
	\begin{align}
	     f_{k\sigma}^{\text{eq}} \sim f_0(E_{k\sigma}) - \sigma \frac{\partial f_0(E_{k\sigma})}{\partial E_{k\sigma}} \delta \mu_s ,
	\end{align}
	where the second term represents the nonequilibrium spin imbalance due to SVC.
	
\begin{figure}
        \centering
        \includegraphics[width=60mm]{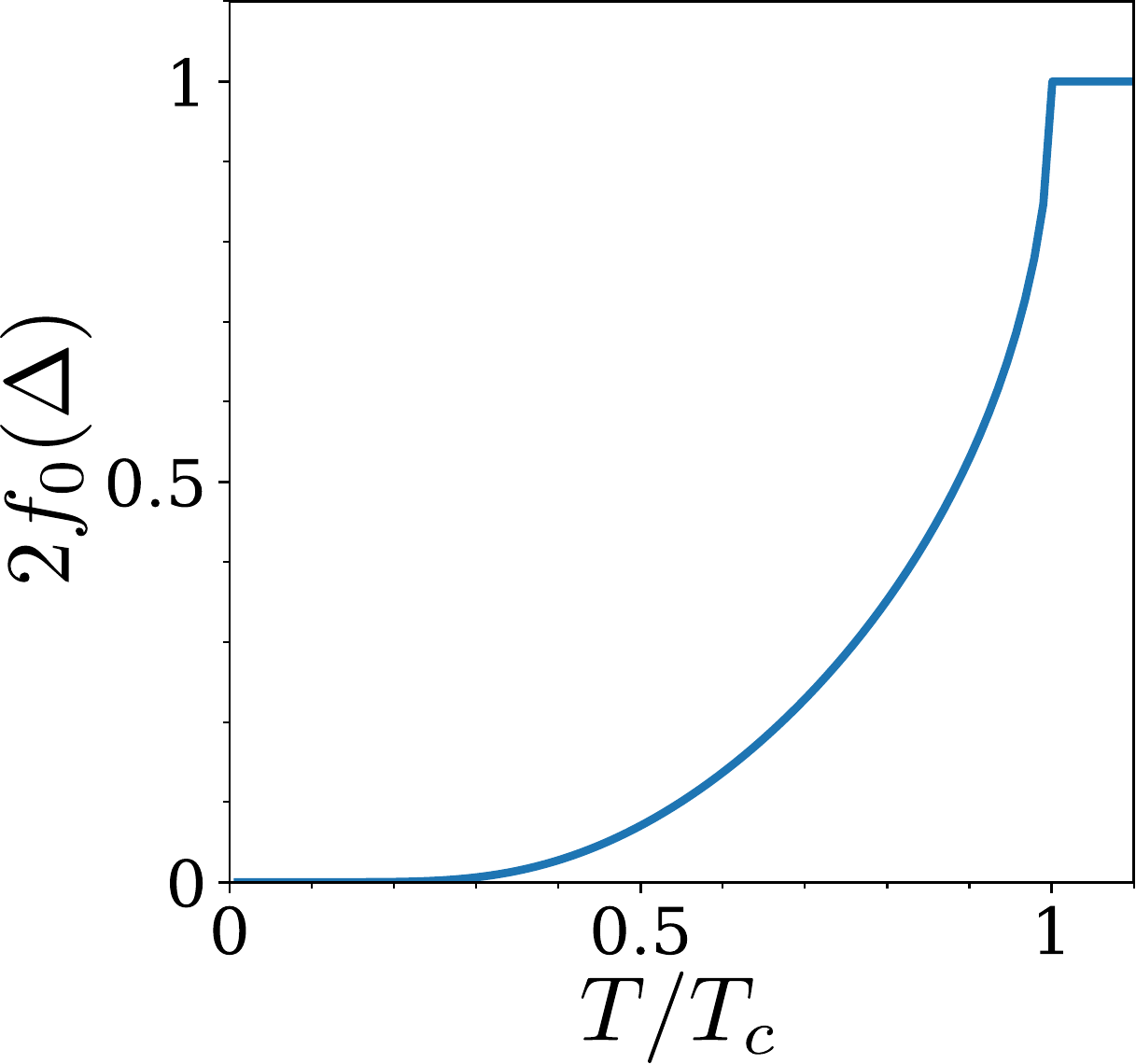}
        \caption{Temperature dependence of the factor $2f_0(\Delta)$.}
        \label{fig:fermi}
    \end{figure}

    It is convenient to introduce the following expansion of the nonequilibrium distribution function:
    \begin{align}
	    f_{\bm k\sigma} = f_{k\sigma}^{\text{eq}} +   f^{\text{diff}}_{\bm k\sigma} + f^{\text{drift}}_{\bm k\sigma}.
	    \label{distribution}
	\end{align}
    Assuming that the spin relaxation time is much longer than the momentum relaxation time, i.e., $\tau_k\ll \tau_{s,k}$, which is well satisfied in metals, the second and third terms in Eq.~(\ref{distribution}) are determined by
    \begin{gather}
          \bm v_{\bm k} \cdot \frac{\partial f_{k\sigma}^{\text{eq}}}{\partial \bm r} = -\frac{f_{\bm k\sigma}^{\text{diff}}}{\tau_k},
          \\
          \bm F_{\sigma} \cdot \bm v_{\bm k} \frac{\partial f_{ k\sigma}^{\text{eq}}}{\partial E_{k}} = -\frac{ f^{\text{drift}}_{\bm k\sigma}}{\tau_k}
          .
    \end{gather}
    These are solved as 
    \begin{gather}
        f_{\bm k\sigma}^{\text{diff}} = \sigma \tau_k \bm v_{\bm k} \cdot \frac{\partial}{\partial \bm r}  \left[ \frac{\hbar}{4} \Omega + \delta \mu_s \right] \frac{\partial f_0(E_k)}{\partial E_k},
        \\
        f^{\text{drift}}_{\bm k\sigma} = -\sigma \tau_k \frac{\hbar}{4} \frac{\partial \Omega}{\partial \bm r}   
        \frac{\partial f_0(E_k)}{\partial E_k}
        .
    \end{gather}
    As can be seen from the results, the terms depending on the spatial gradient of the vorticity cancel out.
    This corresponds to the cancellation between spin current driven by the spin-dependent force due to the SVC and the diffusion spin current.
    The nonequilibrium distribution function is given by
    \begin{align}
        f_{\bm k\sigma} = f_{k\sigma}^{\text{eq}} + \sigma \tau_k \bm v_{\bm k} \cdot \frac{\partial \delta \mu_s}{\partial \bm r} \frac{\partial f_0(E_k)}{\partial E_k}. 
        \label{total}
    \end{align}
    Substituting the nonequilibrium distribution function into Eq.~(\ref{spin_curent_exp}), we derive 
    \begin{align}
        \average{\bm j_s(\bm r,t)} &= -2 \sum_{\bm k} \tau_k \bm v_{\bm k} \bm v_{\bm k} \cdot \frac{\partial \delta \mu_s}{\partial \bm r} \left( -\frac{\partial f_0(E_k)}{\partial E_k}\right).
    \end{align}
    Therefore, the spin current is given by
    \begin{align}
        \bm j_s(\bm r,t)= -\frac{\sigma_c}{e^2} 2f_0(\Delta)\frac{\partial \delta \mu_s}{\partial \bm r},
    \end{align}
    where $\sigma_c=2e^2N(\epsilon_F)D$ is the Drude conductivity in the normal state with $D=v_F^2\tau/3$ being the diffusion constant, $v_F=\hbar k_F/m$ being the Fermi velocity of the electrons, $k_F$ being the Fermi wavenumber, and $N(\epsilon_F)$ being the density of states per spin at the Fermi level.
    The temperature dependence of the spin current is determined by the factor $2f_0(\Delta)$, which is plotted in Fig.~\ref{fig:fermi}.
    This temperature dependence indicates that the opening of a superconducting gap prevents generation of spin current as the temperature decrease.
    
    The spin accumulation is determined from the spin-diffusion equation.
    We substitute the nonequilibrium distribution function in Eq.~(\ref{total}) into the Boltzmann transport equation, Eq.~(\ref{beq1}), and integrate the wavenumber for the difference between the up-spin and down-spin equations.
    The spin-diffusion equation is given by
    \begin{align}
        \left( \frac{\partial}{\partial t} - D_s(T) \frac{\partial^2}{\partial r^2} + \tau_{\text{sf}}^{-1}(T)  \right) \delta \mu_s = -\frac{\hbar}{4} \dot \Omega - \zeta \frac{\hbar \Omega}{2\tau_{\text{sf}}(T)},
        \label{eq.spin_diff}
    \end{align}
    where the first term in the right-hand side is the spin-source term caused by time-dependent Zeeman splitting due to the effective magnetic field of SVC.
    In Eq.~(\ref{eq.spin_diff}), we introduce a second term to the right-hand side that is the spin-source term caused by transverse fluctuation of vorticity with renormalization factor $\zeta$ which is material dependent\cite{matsuo2017Theory}.
    $\tau_{\text{sf}}(T)$ is the spin relaxation time, given by
    \begin{align}
        \tau_{\text{sf}}(T) = \frac{\chi_s}{2 f_0(\Delta)} \tau_{\text{sf}}^n,
    \end{align}
    where $\tau_{\text{sf}}^n=9\tau/4\lambda_{\text{so}}^2k_F^4$ is the spin relaxation time in the normal state, and $\chi_s$ is the susceptibility of the quasiparticle spin:
    \begin{align}
        \chi_{s} &= 2 \int^{\infty}_{\Delta} dE \frac{E}{\sqrt{E^2-\Delta^2}} \left( -\frac{\partial f_0(E)}{\partial E} \right),
    \end{align}
    Finally, $D_s(T)$ is the spin-diffusion constant in the superconducting state, defined by
    \begin{align}
        D_s(T) = \frac{2 f_0(\Delta)}{\chi_s} D.
    \end{align}
    It is noted that the spin-diffusion length in the superconducting state $\lambda_{\text{sf}}=\sqrt{\tau_{\text{sf}}(T)D_s(T)}$ is same as that in the normal state $\lambda_{\text{sf}}=\sqrt{\tau_{\text{sf}}^nD}$.
    Therefore, the spin-source term due to SVC arises in the spin-diffusion equation of the superconductor, which can generate a spin current in the superconductor.

    \begin{figure}
        \centering
        \includegraphics[width=70mm]{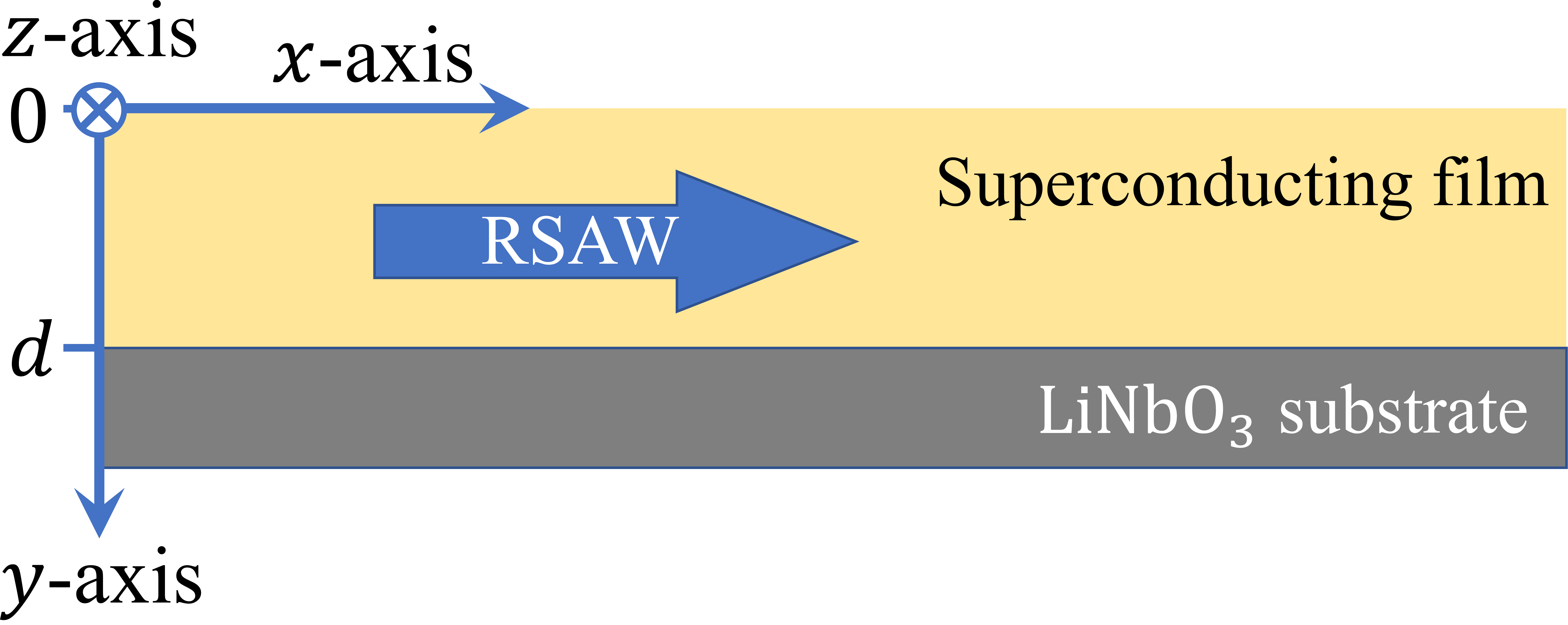}
        \caption{Schematic illustration of an s-wave superconductor film with thickness $d$ to which an Rayleigh-type surface acoustic wave (SAW) is applied.
        }
        \label{fig:film}
    \end{figure}

    \section{Discussion}
        
    Let us consider that a RSAW is applied to an s-wave superconductor thin film with thickness $d$, as shown in Fig.~\ref{fig:film}.
    We choose the $x$ axis as the direction of propagation of the RSAW and the $y$ axis as the depth direction.
    The vorticity associated with the RSAW is oriented along the $z$ axis.
    We assume that the time and spatial variations of the RSAW are much slower than the spin relaxation time and spin-diffusion length in the s-wave superconductor, respectively.
    Therefore, the time and $x$-directional spatial variation of spin accumulation are approximately proportional to the vorticity.
    Additionally, previous studies have suggested that the second term of the spin source term in the right-hand side of Eq.~(\ref{eq.spin_diff}) mainly contributes to generating spin current by the SVC and that the first term of the spin-source term is negligible.
    Hence, we should solve the following one-dimensional stationary spin-diffusion equation:
    \begin{align}
        \left( -\lambda_{\text{sf}}^2\frac{\partial^2}{\partial y^2} + 1
        \right) \delta \mu_s(y) = -\zeta \frac{\hbar \Omega}{2}.
    \end{align}
    The boundary conditions require that no spin current flow across the surfaces, i.e., $\bm j_s(d)=\bm 0$ and $\bm j_s(0)=\bm 0$.
    The lattice displacement due to the RSAW with the wavenumber $q(>0)$ and the frequency $\omega$ is given by
    \begin{align}
        \bm u(\bm r,t) = u_0 e^{i(qx-\omega t)}
        \begin{pmatrix}
            i \left( 
                e^{-\kappa_ly} - \frac{2\kappa_t \kappa_l}{\kappa_t^2+q^2} e^{-\kappa_ty}
            \right)
            \\[10pt]
            \frac{\kappa_l}{|q|}
            \left( 
                -e^{-\kappa_ly} + \frac{2q^2}{\kappa_t^2+q^2} e^{-\kappa_ty}
            \right)
            \\[10pt]
            0
        \end{pmatrix}
        ,
    \end{align}
    where $u_0$ is the amplitude of the RSAW, and $\kappa_t=\sqrt{q^2-\omega^2/c_l^2}$ and $\kappa_l=\sqrt{q^2-\omega^2/c_t^2}$ are the decay constants of transverse and longitudinal waves, respectively, with transverse wave velocity $c_t$ and longitudinal wave velocity $c_l$.
    The vorticity associated with the RSAW $\bm \Omega = \nabla \times \bm u$ is oriented along the $z$-axis, and its $z$-component is given by
    \begin{gather}
        \Omega (\bm r,t) = \dfrac{2r_0\omega^2}{c_R}  e^{-\kappa_t y} e^{i(qx-\omega t)}
        ,
    \end{gather}
    where $c_R=\omega/|q|$ is the velocity of the RSAW and $r_0$ is the amplitude along the $y$-direction, given by
    \begin{align}
        r_0 = |u_y (y=0)| =  \frac{-\kappa_t^2+q^2}{\kappa_t^2+q^2} \frac{\kappa_l}{|q|}u_0.
    \end{align}
    
     \begin{figure}
        \centering
        \includegraphics[width=60mm]{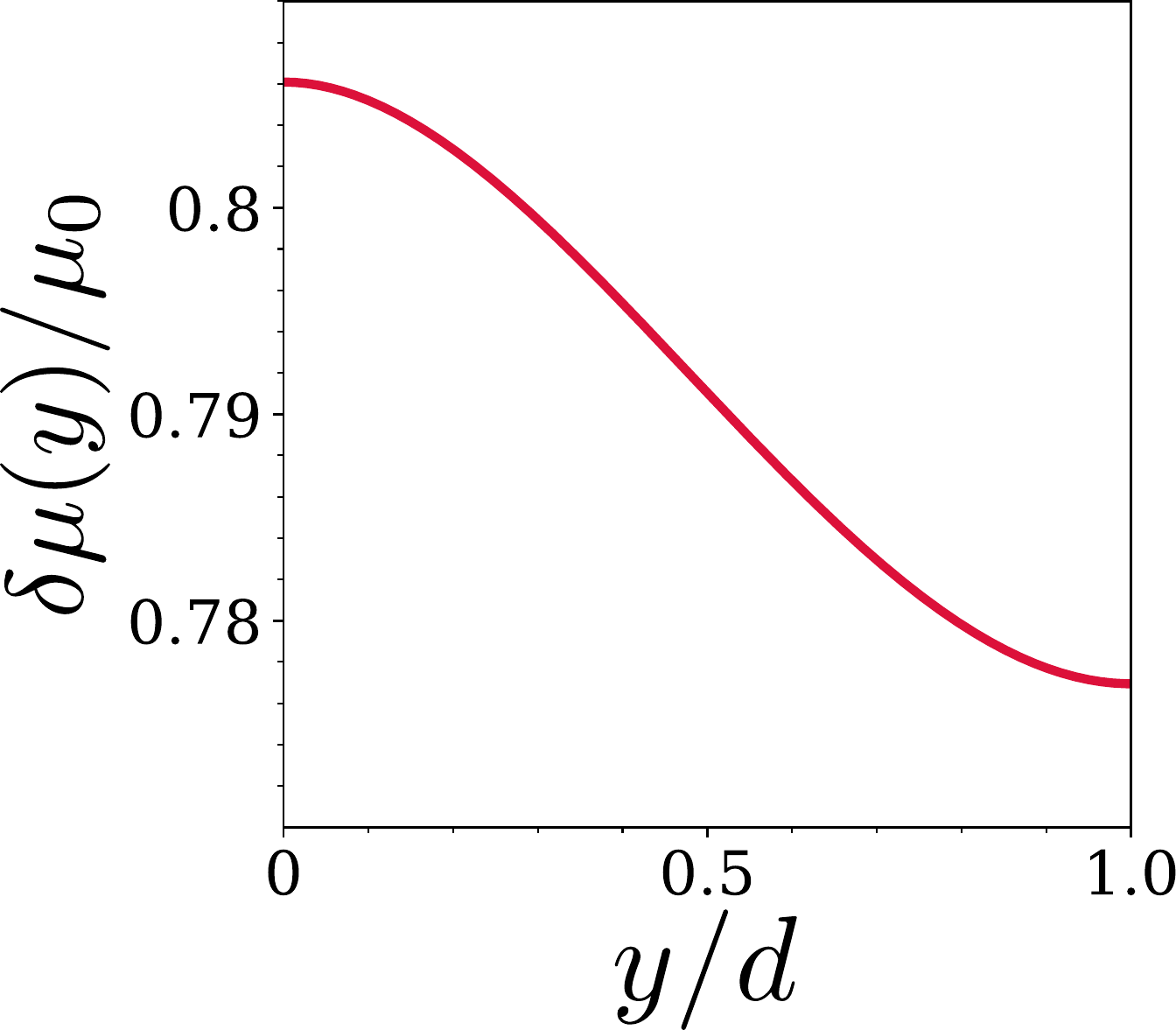}
        \caption{Plot of $y$ dependence of the magnitude of the spin accumulation induced by the RSAW in Al.
        The spin accumulation is normalized by the magnitude of the spin-source term $\mu_0=\hbar \zeta |\Omega (\bm r,t)|_{y=0}/2$.
        The film thickness is assumed to be the same as the spin-diffusion length, $d=\lambda_{\text{sf}}$.
        }
        \label{fig:spin_acc}
    \end{figure}
   
    Solving the spin-diffusion equation under the boundary conditions, the spin accumulation and spin current generated by the RSAW are given by
    \begin{align}
        &\delta \mu_s (y,t) = -\frac{\hbar \zeta |\Omega(\bm r,t)|_{y=0}e^{i(qx-\omega t)}}{2(1-\kappa_t^2\lambda_{\text{sf}}^2)} \biggl\{
        e^{-\kappa_t y}
        \nonumber\\
        &+ \frac{\lambda_{\text{sf}} \kappa_t [ e^{-\kappa_t d}\cosh (y/\lambda_{\text{sf}}) - \cosh ((d-y)/\lambda_{\text{sf}}) ]}{\sinh (d/\lambda_{\text{sf}})}
        \biggr\},
    \\
        &j_{\text s,y}(y,t) = \frac{\sigma_c}{e^2}2f_0(\Delta) \frac{\hbar \zeta |\Omega(\bm r,t)|_{y=0}e^{i(qx-\omega t)}}{2(1-\kappa_t^2\lambda_{\text{sf}}^2)} \biggl\{
        -\kappa_t e^{-\kappa_t y}
        \nonumber \\
        &+ \frac{ \kappa_t [ e^{-\kappa_t d}\sinh (y/\lambda_{\text{sf}}) + \sinh ((d-y)/\lambda_{\text{sf}}) ]}{\sinh (d/\lambda_{\text{sf}})}
        \biggr\},
    \end{align}
    where $|\Omega(\bm r,t)|_{y=0}$ is the magnitude of the vorticity at the surface.
    
    \begin{figure}
        \centering
        \includegraphics[width=70mm]{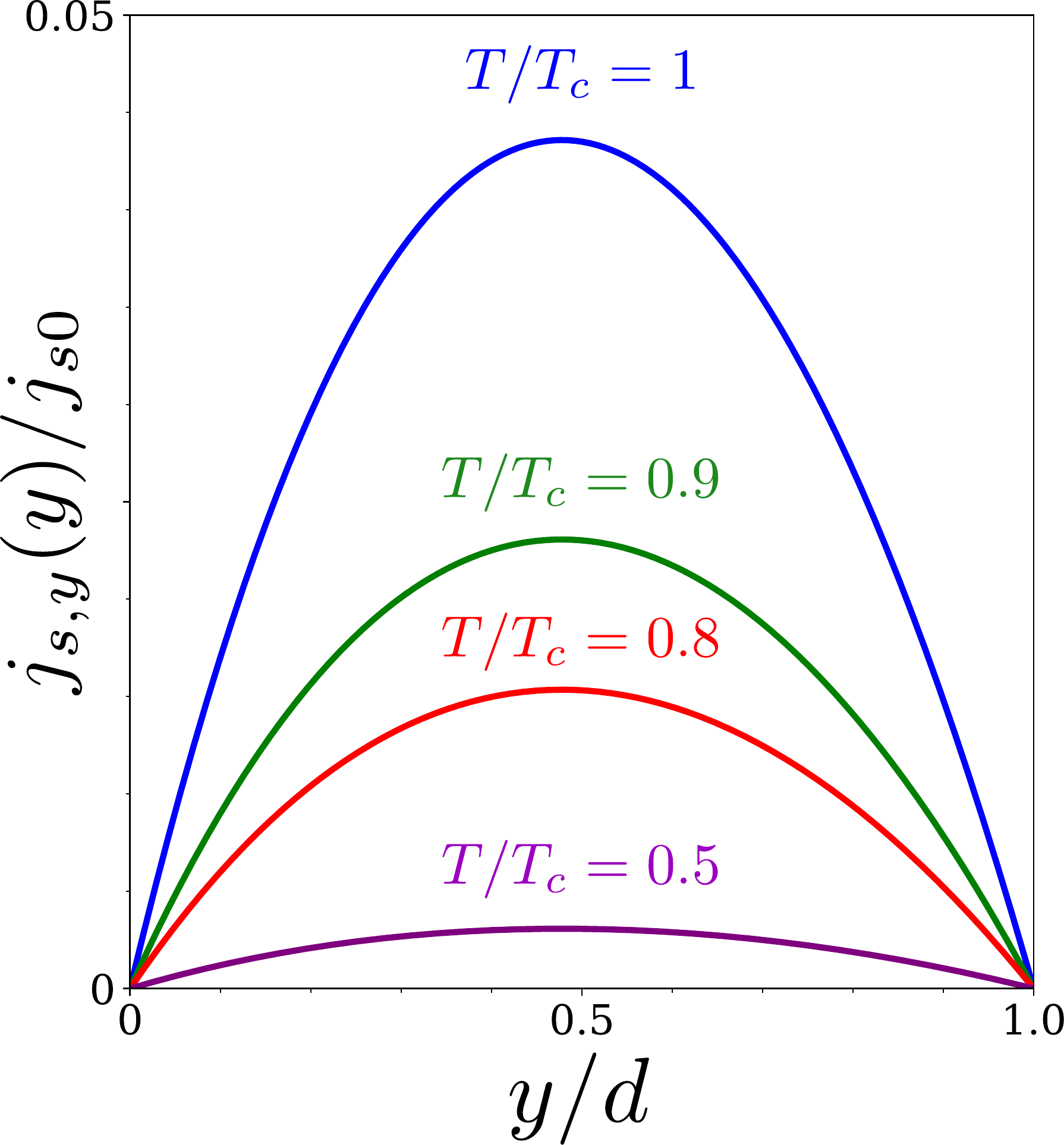}
        \caption{Spin current generated by an RSAW in superconductor Al.
        The spin current is normalized by $j_{s0}=\sigma_c\mu_0/e^2\lambda_{\text{sf}}$.
        Line colors indicate the superconductor temperature.
        }
        \label{fig:js}
    \end{figure}

 Next, we estimate the spin current generated by an RSAW in  superconducting Al with a long spin-diffusion length $\lambda_{\text{sf}}\sim 1\, \mu \text m$.
    We consider that an RSAW with frequency $\omega=10\,\text{GHz}$ is excited on a piezoelectric LiNbO$_3$ substrate with a longitudinal wave velocity $c_l=6.75\times 10^3\,\text{m/s}$, transverse wave velocity  $c_t=4.07 \times 10^3\,\text{m/s}$, RSAW velocity  $c_R = 3.99\times 10^3\,\text{m/s}$, and transverse wave decay constant $\kappa_t = 4.88\times 10^5\,\text{m}^{-1}$.
    The magnitude of the vorticity is calculated as $|\Omega(\bm r,t)|_{y=0}\sim 5.0 \times 10^4\,\text s$ with the amplitude of the lattice displacement $r_0\simeq 10^{-12}\,\text m$.
    Previous work\cite{kurimune2020Highly} proposed that the renormalization factor can be estimated by $\zeta \simeq 10^6$.
    The $y$-dependence of the spin accumulation normalized by $\mu_0$ is plotted in Fig.~\ref{fig:spin_acc}, and the spin-current profiles are plotted in Fig.~\ref{fig:js}.
    According to the results, the spin accumulation is independent of the temperature even in a superconducting state.
    Conversely, the spin current generated by the RSAW strongly depends on the temperature.
    This suggests that the superconducting gap opens as the temperature decreases, and the generation of spin current is suppressed.
    Here, the magnitude of the spin-source term is estimated as
    $\mu_0 \equiv \frac{\zeta}{2} \hbar |\Omega(\bm r,t)|_{y=0} \simeq 1.65 \times 10^{-5}\, \text{eV}$, and the magnitude of the spin current is estimated as
    $ej_{\text s0} \equiv \sigma_c \mu_0/e\lambda_{\text{sf}} \simeq 2.8 \times 10^8\, \text A\cdot \text m^{-2}$, where $\sigma_c=1.7\times 10^7\, \Omega^{-1}m^{-1}$.
    The detectable spin current can be generated in a single-superconductor film.

    \section{conclusion}
    
    We theoretically studied spin-current generation in an s-wave superconductor by a SAW via SVC.
    The spin-diffusion equation, for which the spin accumulation of the quasiparticle satisfies, has been derived up to the first order in vorticity based on kinetic theory.
    Using the results, we calculated the spin current generated by the SAW in superconducting Al.
    The results suggest that spin transport with quasiparticles can be driven by mechanical means in a single-superconductor layer.
    It is found that the generation of spin current in superconductors is suppressed since the superconducting gap opens when the temperature is low.
    Our estimation suggests that an observable magnitude of spin current can be induced.
    Our prediction may provide support for the development of spin transport in superconductors.

    \begin{acknowledgments}
    The authors would like to greatly acknowledge the continued support of Y. Nozaki.
    We also thank T. Horaguchi and H, Nakayama for daily discussions.
    The authors would like to thank MARUZEN-YUSHODO Co., Ltd. ( https://kw.maruzen.co.jp/kousei-honyaku/ ) for the English language editing.
	This work was partially supported by JST CREST Grant No. JPMJCR19J4, Japan.
	This work was supported by JSPS KAKENHI for Grant Nos. 20H01863, 20K03831, 21H04565, 21H01800, and 21K20356.
	MM was supported by the Priority Program of the Chinese Academy of Sciences, Grant No. XDB28000000.
	\end{acknowledgments}
	

%

	\end{document}